\documentclass[12pt]{article}
\usepackage{epsfig}
\usepackage{amsmath}
\usepackage{hhline}
\usepackage{amssymb}
\usepackage{times}
\usepackage{cite}

\newlength{\dinwidth}
\newlength{\dinmargin}
\begin{document}  
\newcommand{\pom}{{I\!\!P}}
\newcommand{\reg}{{I\!\!R}}
\newcommand{\slowpi}{\pi_{\mathit{slow}}}
\newcommand{\fiidiii}{F_2^{D(3)}}
\newcommand{\fiidiiiarg}{\fiidiii\,(\beta,\,Q^2,\,x)}
\newcommand{\n}{1.19\pm 0.06 (stat.) \pm0.07 (syst.)}
\newcommand{\nz}{1.30\pm 0.08 (stat.)^{+0.08}_{-0.14} (syst.)}
\newcommand{\fiidiiiful}{F_2^{D(4)}\,(\beta,\,Q^2,\,x,\,t)}
\newcommand{\fiipom}{\tilde F_2^D}
\newcommand{\ALPHA}{1.10\pm0.03 (stat.) \pm0.04 (syst.)}
\newcommand{\ALPHAZ}{1.15\pm0.04 (stat.)^{+0.04}_{-0.07} (syst.)}
\newcommand{\fiipomarg}{\fiipom\,(\beta,\,Q^2)}
\newcommand{\pomflux}{f_{\pom / p}}
\newcommand{\nxpom}{1.19\pm 0.06 (stat.) \pm0.07 (syst.)}
\newcommand {\gapprox}
   {\raisebox{-0.7ex}{$\stackrel {\textstyle>}{\sim}$}}
\newcommand {\lapprox}
   {\raisebox{-0.7ex}{$\stackrel {\textstyle<}{\sim}$}}
\def\gsim{\,\lower.25ex\hbox{$\scriptstyle\sim$}\kern-1.30ex%
\raise 0.55ex\hbox{$\scriptstyle >$}\,}
\def\lsim{\,\lower.25ex\hbox{$\scriptstyle\sim$}\kern-1.30ex%
\raise 0.55ex\hbox{$\scriptstyle <$}\,}
\newcommand{\pomfluxarg}{f_{\pom / p}\,(x_\pom)}
\newcommand{\dsf}{\mbox{$F_2^{D(3)}$}}
\newcommand{\dsfva}{\mbox{$F_2^{D(3)}(\beta,Q^2,x_{I\!\!P})$}}
\newcommand{\dsfvb}{\mbox{$F_2^{D(3)}(\beta,Q^2,x)$}}
\newcommand{\dsfpom}{$F_2^{I\!\!P}$}
\newcommand{\gap}{\stackrel{>}{\sim}}
\newcommand{\lap}{\stackrel{<}{\sim}}
\newcommand{\fem}{$F_2^{em}$}
\newcommand{\tsnmp}{$\tilde{\sigma}_{NC}(e^{\mp})$}
\newcommand{\tsnm}{$\tilde{\sigma}_{NC}(e^-)$}
\newcommand{\tsnp}{$\tilde{\sigma}_{NC}(e^+)$}
\newcommand{\st}{$\star$}
\newcommand{\sst}{$\star \star$}
\newcommand{\ssst}{$\star \star \star$}
\newcommand{\sssst}{$\star \star \star \star$}
\newcommand{\tw}{\theta_W}
\newcommand{\sw}{\sin{\theta_W}}
\newcommand{\cw}{\cos{\theta_W}}
\newcommand{\sww}{\sin^2{\theta_W}}
\newcommand{\cww}{\cos^2{\theta_W}}
\newcommand{\trm}{m_{\perp}}
\newcommand{\trp}{p_{\perp}}
\newcommand{\trmm}{m_{\perp}^2}
\newcommand{\trpp}{p_{\perp}^2}
\newcommand{\alp}{\alpha_s}

\newcommand{\alps}{\alpha_s}
\newcommand{\sqrts}{$\sqrt{s}$}
\newcommand{\LO}{$O(\alpha_s^0)$}
\newcommand{\Oa}{$O(\alpha_s)$}
\newcommand{\Oaa}{$O(\alpha_s^2)$}
\newcommand{\PT}{p_{\perp}}
\newcommand{\JPSI}{J/\psi}
\newcommand{\sh}{\hat{s}}
\newcommand{\uh}{\hat{u}}
\newcommand{\MP}{m_{J/\psi}}
\newcommand{\PO}{I\!\!P}
\newcommand{\xbj}{x}
\newcommand{\xpom}{x_{\PO}}
\newcommand{\ttbs}{\char'134}
\newcommand{\xpomlo}{3\times10^{-4}}  
\newcommand{\xpomup}{0.05}  
\newcommand{\dgr}{^\circ}
\newcommand{\pbarnt}{\,\mbox{{\rm pb$^{-1}$}}}
\newcommand{\gev}{\,\mbox{GeV}}
\newcommand{\WBoson}{\mbox{$W$}}
\newcommand{\fbarn}{\,\mbox{{\rm fb}}}
\newcommand{\fbarnt}{\,\mbox{{\rm fb$^{-1}$}}}
\newcommand{\dsdx}[1]{$d\sigma\!/\!d #1\,$}
\newcommand{\eV}{\mbox{e\hspace{-0.08em}V}}
%
%
\newcommand{\qsq}{\ensuremath{Q^2} }
\newcommand{\gevsq}{\ensuremath{\mathrm{GeV}^2} }
\newcommand{\et}{\ensuremath{E_t^*} }
\newcommand{\rap}{\ensuremath{\eta^*} }
\newcommand{\gp}{\ensuremath{\gamma^*}p }
\newcommand{\dsiget}{\ensuremath{{\rm d}\sigma_{ep}/{\rm d}E_t^*} }
\newcommand{\dsigrap}{\ensuremath{{\rm d}\sigma_{ep}/{\rm d}\eta^*} }

\newcommand{\dstar}{\ensuremath{D^*}}
\newcommand{\dstarp}{\ensuremath{D^{*+}}}
\newcommand{\dstarm}{\ensuremath{D^{*-}}}
\newcommand{\dstarpm}{\ensuremath{D^{*\pm}}}
\newcommand{\zDs}{\ensuremath{z(\dstar )}}
\newcommand{\Wgp}{\ensuremath{W_{\gamma p}}}
\newcommand{\ptds}{\ensuremath{p_t(\dstar )}}
\newcommand{\etads}{\ensuremath{\eta(\dstar )}}
\newcommand{\ptj}{\ensuremath{p_t(\mbox{jet})}}
\newcommand{\ptjn}[1]{\ensuremath{p_t(\mbox{jet$_{#1}$})}}
\newcommand{\etaj}{\ensuremath{\eta(\mbox{jet})}}
\newcommand{\detadsj}{\ensuremath{\eta(\dstar )\, \mbox{-}\, \etaj}}

\def\Journal#1#2#3#4{{#1} {\bf #2} (#3) #4}
\def\NCA{\em Nuovo Cimento}
\def\NIM{\em Nucl. Instrum. Methods}
\def\NIMA{{\em Nucl. Instrum. Methods} {\bf A}}
\def\NPB{{\em Nucl. Phys.}   {\bf B}}
\def\PLB{{\em Phys. Lett.}   {\bf B}}
\def\PRL{\em Phys. Rev. Lett.}
\def\PRD{{\em Phys. Rev.}    {\bf D}}
\def\ZPC{{\em Z. Phys.}      {\bf C}}
\def\EJC{{\em Eur. Phys. J.} {\bf C}}
\def\CPC{\em Comp. Phys. Commun.}

\begin{titlepage}

\noindent
\begin{flushleft}
{\tt DESY 11-183    \hfill    ISSN 0418-9833} \\
{\tt October 2011}                  \\
\end{flushleft}

\noindent

\vspace{2cm}
\begin{center}
\begin{Large}

{\bf Measurement of the Azimuthal Correlation \\
between the most Forward Jet and the Scattered Positron \\
in Deep-Inelastic Scattering at HERA }

\vspace{2cm}

H1 Collaboration

\end{Large}
\end{center}

\vspace{2cm}

\begin{abstract}
Deep-inelastic positron-proton scattering events at low photon virtuality $Q^2$ with a forward jet, produced at small angles with respect to the proton beam, are measured with the H1 detector at HERA. 
A subsample of events with an additional jet in the 
central region is also studied. 
For both samples differential cross sections and normalised distributions are measured
as a function of the azimuthal angle difference, $\Delta \phi$, between the forward jet and the scattered positron.
The sensitivity to QCD evolution mechanisms is tested by comparing the data to 
predictions of Monte Carlo generators based on different evolution approaches as well as to next-to-leading order calculations.
\end{abstract}

\vspace{1.5cm}

\begin{center}
Submitted to Eur. Phys. J. {\bf C}
\end{center}

\end{titlepage}

%
%
%
\begin{flushleft}

F.D.~Aaron$^{5,48}$,           
C.~Alexa$^{5}$,                
V.~Andreev$^{25}$,             
S.~Backovic$^{30}$,            
A.~Baghdasaryan$^{38}$,        
S.~Baghdasaryan$^{38}$,        
E.~Barrelet$^{29}$,            
W.~Bartel$^{11}$,              
K.~Begzsuren$^{35}$,           
A.~Belousov$^{25}$
% ...

\bigskip{\it
% ...
$ ^{5}$ National Institute for Physics and Nuclear Engineering (NIPNE) ,
          Bucharest, Romania$^{ m}$ \\
% ...
$ ^{11}$ DESY, Hamburg, Germany \\
% ...
$ ^{25}$ Lebedev Physical Institute, Moscow, Russia$^{ e}$ \\
% ...
$ ^{29}$ LPNHE, Universit\'e Pierre et Marie Curie Paris 6,
          Universit\'e Denis Diderot Paris 7, CNRS/IN2P3, Paris, France \\
 $ ^{30}$ Faculty of Science, University of Montenegro,
          Podgorica, Montenegro$^{ n}$ \\
% ...
$ ^{35}$ Institute of Physics and Technology of the Mongolian
          Academy of Sciences, Ulaanbaatar, Mongolia \\
% ...
$ ^{38}$ Yerevan Physics Institute, Yerevan, Armenia \\
% ...
$ ^{48}$ Also at Faculty of Physics, University of Bucharest,
          Bucharest, Romania \\
% ...
$ ^e$ Supported by the Deutsche Forschungsgemeinschaft \\
% ...
$ ^m$ Supported by the Romanian National Authority for Scientific Research
      under the contract PN 09370101 \\
 $ ^n$ Partially Supported by Ministry of Science of Montenegro,
      no. 05-1/3-3352 \\
}
\end{flushleft}
%

\newpage
\section{Introduction}
Measurements of the hadronic final state in deep-inelastic lepton-proton scattering (DIS) test
Quantum Chromodynamics (QCD), the theory of the strong force. 
At moderate negative four-momentum 
transfers squared $Q^2$  of a few ${\rm GeV}^2$, 
the HERA $ep$ collider has extended the available kinematic range for deep-inelastic scattering  to regions of small 
Bjorken-$x\simeq 10^{-4}$.
This is the region of high parton densities in the proton, dominated by gluons and
sea quarks. At the large $\gamma^*p$ centre-of-mass energy available at small $x$, a transition is expected from 
parton cascades ordered in transverse momentum, described by the 
Dokshitzer-Gribov-Lipatov-Altarelli-Parisi (DGLAP) evolution equations \cite{DGLAP},  
to cascades unordered in transverse momentum, described by the Balitsky-Fadin-Kuraev-Lipatov (BFKL) approach \cite{BFKL}.

\begin{figure}[hhh]
\center
\epsfig{file=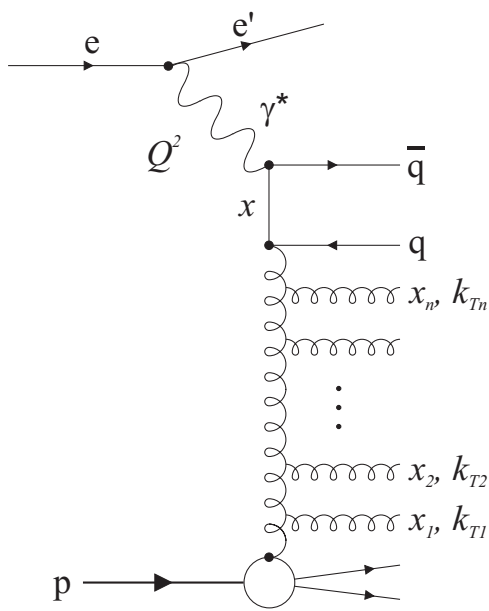 ,width=7cm}
\setlength{\unitlength}{1cm}
\caption{Generic diagram for deep-inelastic $ep$ scattering at small $x$. A
gluon cascade 
evolves between the quark box, attached to the virtual photon, and the proton. The gluon longitudinal 
momentum fractions and transverse momenta are labeled $x_i$ and $k_{Ti}$, respectively.} 
\label{fig:diagram} 
\end{figure}

A generic diagram for
parton evolution in a DIS process at low $x$, in which a gluon from the proton induces a QCD 
cascade before an interaction with the virtual photon, is shown in figure~\ref{fig:diagram}.
In the DGLAP approximation the struck quark originates  from a parton cascade ordered in 
virtualities of the propagator partons. At low $x$ this implies a strong ordering in transverse 
momentum, $k_T$, of the emitted partons, measured with respect to the proton direction.
In the  BFKL  approach  there is no ordering in $k_T$ of the partons 
along the ladder. Compared to the DGLAP scheme more gluons with sizable transverse 
momentum are emitted near the proton direction. 
For this reason energetic jets of high transverse momentum produced close to the proton direction in the laboratory frame, referred 
to as the forward region, are considered to be especially sensitive to QCD
dynamics at low $x$~\cite{Mueller}. 
Forward jet production was measured previously by the H1 and ZEUS collaborations.
In these measurements as well as in the  present one, the
requirements on the forward jet and the phase space were  chosen in such a way that the standard 
DGLAP evolution is suppressed and the effects of BFKL dynamics are enhanced. Preference 
for models which employ QCD evolution non-ordered in transverse momentum was 
observed~\cite{h1fj00,h1fj01,h1fj02,zeusfj01,zeusfj02,zeusfj03,zeusfj04}. 

One of the observables suggested to be
sensitive to BFKL dynamics~\cite{Bartels} is the azimuthal angle 
difference, $\Delta \phi$, between the forward jet and 
the scattered electron, defined in the laboratory frame. 
In the Quark Parton Model (QPM) process  $e+q \rightarrow e+q$ the simple two-body kinematics constrains the scattered electron 
and the jet to be produced back-to-back, and thus predicts at the parton level $\Delta \phi = \pi$.  Hadronisation effects induce some smearing to this parton level prediction. 
Inclusion of higher order processes  partially decorrelates the jet from the electron. 
As a consequence, 
for evolution schemes without ordering  in transverse momentum, the decorrelation is expected 
to increase with  electron-jet rapidity distance, $Y$, 
since the phase space  for additional parton emissions increases.    
The calculations  employing the BFKL approach to the next-to-leading 
order accuracy (NLO BFKL), indeed predict an increase of the azimuthal angle decorrelation with 
the electron-jet rapidity distance \cite{NLOBFKL}. 

This paper presents a study of low $x$ DIS interactions in which high transverse momentum jets 
are produced in the forward region.
The forward jet cross sections and normalised distributions are measured
as a function of the azimuthal angle difference $\Delta \phi$  in three bins
of the rapidity separation $Y$ between the positron and the forward jet. The forward jet cross section 
as a function of  $Y$ is also measured.
Moreover, the measurements of the azimuthal correlations  in $\Delta \phi$ are performed 
using a subsample defined by a requirement of 
an additional central jet. In comparison with the forward jet  sample, this  
subsample is expected to contain a higher fraction of forward jets 
from additional gluon emissions.

The data set used for the analysis was collected with the H1 detector in the
year $2000$,  when 
positrons and  protons collided with energies of $27.6$ ${\rm GeV}$
and $920$ ${\rm GeV}$, respectively, corresponding to a centre-of-mass energy of $\sqrt{s} = 319$ ${\rm GeV}$. 
The integrated luminosity of 
the data set is $38.2$ ${\rm pb}^{-1}$, which is 
about fourteen times larger than in the previous
measurement of the azimuthal decorrelation of forward jets \cite{h1fj01}. 

\section{QCD Calculations}
The measurements presented here are compared with predictions of 
Monte Carlo (MC) programs and perturbative QCD calculations at 
next-to-leading order (NLO).
The MC programs  use first-order QCD matrix elements 
and model higher order terms by parton showers in the leading logarithm approximation 
or by quasi-classical gluon radiation from colour dipoles.  
Three MC event generators, which adopt different QCD based approaches to model 
the parton cascade, are used. 
\begin{itemize}
\item
RAPGAP \cite{RAPGAP}  matches first order QCD matrix elements to DGLAP based  leading-log parton showers with $k_T$ ordering.
The factorisation and renormalisation scales are set to
$\mu_f =  \mu_r = \sqrt{Q^2 + p^2_{\rm T}}$, where $p_{\rm T}$ is the transverse momentum of the two outgoing hard partons in 
the centre-of-mass of the hard subsystem.
Predictions of RAPGAP are labeled DGLAP in  the figures. 
\item
DJANGOH \cite{DJANGOH} with ARIADNE includes an implementation of the Colour Dipole Model (CDM) \cite{CDM}, which has as its basic construct a colour dipole formed by the struck quark and the proton remnant. Subsequent parton emissions originate from a chain of independently radiating dipoles formed by the emitted gluons.
In this approach the transverse momenta of emitted gluons perform  a random walk such that CDM provides a BFKL-like approach. The leading order partonic final state is corrected 
to exactly reproduce the $O(\alpha_S)$ matrix elements.
The simulation
of DJANGOH/ARIADNE uses a set of colour dipole parameters tuned to describe measurements 
of the hadronic final state in DIS at HERA \cite{thesis}.
The DJANGOH/ARIADNE predictions are referred to as CDM in the following. 
\item
CASCADE \cite{CASCADE} implements the Ciafaloni-Catani-Fiorani-Marchesini (CCFM) 
evolution \cite{CCFM} which aims to unify the DGLAP and BFKL approaches. It introduces 
angular ordering of emissions to implement gluon coherence effects, and thus 
in the high energy limit  the CCFM evolution equation is almost equivalent to the BFKL 
approach, while reproducing the DGLAP equations for large $x$ and high $Q^2$. CASCADE 
uses off-shell leading order QCD matrix elements, supplemented with gluon emissions based 
on the CCFM evolution equation, requiring an unintegrated gluon density function (uPDF) which takes the 
transverse momenta of the propagators into account.
In this paper two different uPDF sets are used: 
set A0 \cite{setA0}  with only singular terms of the gluon splitting function
and J2003-set 2 \cite{set2} including also non-singular terms, labeled set 2 in the figures.
These parameterisations for the  unintegrated gluon density were obtained using the CCFM 
evolution equation to describe the structure function $F_2(x,Q^2)$  as measured 
by H1 \cite{F2H1} and ZEUS \cite{F2ZEUS}.
Predictions of CASCADE are labeled CCFM in the figures.
\end{itemize}
To perform the hadronisation step, all of the above models use the Lund string fragmentation 
scheme, as implemented in JETSET \cite{JETSET1} in the case of DJANGOH/ARIADNE and 
in PYTHIA~\cite{PYTHIA} for RAPGAP and CASCADE, using a tuning based on LEP 
$e^+e^-$ data \cite{ALEPH}. 
The RAPGAP and DJANGOH/ARIADNE predictions are calculated using the HERAPDF1.0 
\cite{HERAPDF} set of parton distribution functions (PDF).  

The RAPGAP and DJANGOH/ARIADNE programs are interfaced with 
HERACLES~\cite{HERACLES}, which allows the simulation of 
QED-radiative effects. 
These MC models are used to simulate detector effects in order to
determine the acceptance and efficiency for selected forward jet events in DIS.
Generated events are passed through a GEANT~\cite{GEANT} based simulation
of the H1 apparatus,  which takes into account the running conditions of the 
data taking.
Simulated events are reconstructed and analysed using the same program chain 
as is used for the data.

The measurements of azimuthal 
correlations are also compared to the fixed order NLO DGLAP 
predictions of NLOJET++~\cite{NLOJET++}. 
The NLOJET++ program is used here to calculate dijet production
at parton level in DIS at NLO($\alpha_S^2$) accuracy. 
It should be noted that the jet search is performed on partons in the Breit frame (see section $3.2$), and therefore 
the events contain at least one jet in addition to the forward jet.
The renormalisation and factorisation 
scales are defined for each event and are set to
$\mu_r = \mu_f = \sqrt{(P_{\rm T,sc}^2 + Q^2) / 2}$, 
where $P_{\rm T,sc}$ is the transverse momentum 
of the forward jet or the average transverse momentum of the forward and central jet 
in the forward jet sample and in the sample with an additional central jet, respectively. 
The NLO calculations are performed using  the 
CTEQ6.6 \cite{CTEQ6.6} parameterisation of the parton distributions in the proton. 
The NLOJET++ parton level cross sections are corrected for hadronisation effects 
using the RAPGAP model. 
The correction factors for hadronisation are estimated bin-by-bin by calculating the 
ratio between the cross section for jets reconstructed
from stable hadrons (hadron level)  and the parton level cross section. 
The correction factors for hadronisation are in the range from $0.90$ to $1.08$, increasing with 
rapidity distance $Y$. 
The uncertainty of the NLOJET++ predictions due to
missing higher orders is estimated by applying a factor $2$ or $1/2$ to the 
renormalisation and factorisation scales simultaneously.
\section{Experimental Method}
\subsection{H1 detector}
A detailed description of the H1 detector can be found elsewhere\cite{h1det01,h1det02,h1det03}.
The components of the detector which are most relevant for this analysis are briefly described below.
The origin of the H1 coordinate system is the nominal $ep$ interaction point. The  direction 
of the proton beam 
defines the positive $z$-axis. Transverse momenta $p_{\rm T}$ and  polar angles $\theta$ of all 
particles are defined with respect to 
this direction. The azimuthal angle $\phi$ defines the particle direction in the transverse plane. 
The pseudorapidity is given by $\eta =-$ln (tan $\theta$/2).

The $ep$ interaction region is surrounded by the central tracking detector (CTD) consisting of two large concentric drift chambers, 
operated inside a $1.16$ $\rm T$ solenoidal magnetic field.
Charged particles are measured in the angular range
$20^\circ < \theta < 160^\circ$  with a transverse momentum resolution of  
$\sigma_{p_{\rm T}}/{p_{\rm T}} \approx 0.005\cdot p_{\rm T} [\rm GeV] \oplus 0.015$. 
Information from the CTD is used to  trigger events, to 
locate the event vertex,  and contributes to the reconstruction of the hadronic final state. 
 
A highly segmented liquid argon (LAr) calorimeter is used to measure 
the hadronic final state. It covers the range of 
the polar angle $4^\circ < \theta < 154^\circ$ 
and offers full azimuthal coverage. 
The LAr calorimeter consists of an electromagnetic section with lead absorbers and a hadronic section with steel absorbers. 
The total depth of both sections varies between $4.5$ and $8$ interaction lengths in the
region $4^\circ < \theta < 128^\circ$, and between $20$ and $30$ radiation lengths in the region
$4^\circ < \theta < 154^\circ$ increasing towards the forward direction. 
Test beam measurements of the LAr calorimeter modules
showed an energy resolution of 
$\sigma_E/E  \approx 0.50/\sqrt{E[\rm GeV]} \oplus 0.02$ 
for charged pions \cite{h1det04} and of $\sigma_E/E \approx 0.12/\sqrt{E[\rm GeV]} \oplus 0.01$ for
electrons \cite{h1det05}.

A lead/scintillating fiber calorimeter (SpaCal) \cite{h1det03} covers the region 
$153^\circ < \theta < 177.5^\circ$. It has  an electromagnetic and a hadronic section and  is used to 
measure the scattered positron and the backward hadronic energy flow. 
The energy resolution, determined from test beam measurements \cite{h1det06}, 
is $\sigma_E/E  \approx 0.07/\sqrt{E [\rm GeV] } \oplus 0.01$  for electrons. 
The precision of the measurement  of the polar angle of the positron, 
 improved using the backward drift chamber (BDC) situated in front of the SpaCal calorimeter, is
 $1$ $\rm mrad$.    

The luminosity determination is based on the measurement of the Bethe-Heitler process 
$ep \rightarrow ep\gamma$ where the photon
is detected in a calorimeter located at $z = -103$ ${\rm m}$ downstream of the interaction region 
in the positron beam direction.

\subsection{Event selection}
DIS events are selected using  triggers based on electromagnetic energy deposits in the SpaCal calorimeter and 
the presence of charged particle tracks in the central tracker. The trigger efficiency is determined using independently triggered data.
For DIS events with a forward jet, the trigger efficiency lies 
between $60\%$ and $80\%$, and for the topology 'forward and  central jet' it is 
at the level of $80\%$.   
 
The data set is restricted in inelasticity $y$, photon virtuality $Q^2$ 
and $x$: $0.1 < y < 0.7$, $5 < Q^2 < 85$ ${\rm GeV}^2$, $0.0001 < x < 0.004$.  
In this analysis these variables are determined from measurements of the scattered positron energy
and its polar angle, and from the incident positron beam energy.
This phase space is chosen to ensure that the DIS kinematics are well determined and to  
reduce the background from photoproduction.

The background from photoproduction and from events with large initial-state QED radiation is further reduced
by requiring 
$35 < \Sigma_i(E_i - p_{z,i}) < 70$ $\rm GeV$. Here $E_i$ and $p_{z,i}$ are the energy and 
longitudinal momentum of a particle $i$, respectively, and the sum extends over all 
detected particles in the event. 
Energy-momentum conservation requires that
$\Sigma_i(E_i - p_{z,i}) = 2 \cdot E^{0}_{e}$, where $E^{0}_{e}$ is the
positron beam energy.
Jets are identified from combined calorimeter and track objects \cite{object} using the $k_T$ cluster algorithm 
in the longitudinally invariant inclusive mode~\cite{ktalgo} applied in the Breit frame.
The reconstructed jets are then boosted to the laboratory frame.

The measurements of forward jets are restricted to the phase space region where the transverse momentum of the jet is approximately equal to the photon virtuality, $P_{\rm T,fwdjet}^2 \approx Q^2$. This condition
suppresses the contribution of $k_T$-ordered DGLAP cascades  with respect to 
processes unordered in $k_T$~\cite{Mueller}. The selection 
of forward jets with a large fraction of the proton  energy, $x_{\rm fwdjet} \equiv E_{\rm fwdjet}/E_p$, 
such that $x_{\rm fwdjet} \gg x$, enhances
the phase space for BFKL evolution with gluon cascades strongly ordered in fractional 
longitudinal momentum. The above conditions are fulfilled by the requirement that the 
analysed sample contains 
at least one forward jet which satisfies the following criteria in the laboratory frame:  
$P_{\rm T,fwdjet} > 6$~${\rm GeV}$, $ 1.73 < \eta_{\rm fwdjet} < 2.79$,
$x_{\rm fwdjet} > 0.035$ and $0.5 < P^2_{\rm T,fwdjet}/Q^2 < 6$.
Here $\eta_{\rm fwdjet}$ is the pseudorapidity of the forward jet.    
If there is more than one jet fulfilling the above requirements, the jet with the largest pseudorapidity
is chosen.   
The upper cut on $P^2_{\rm T,fwdjet}/Q^2$  is chosen so large in order to reduce 
the contributions of 
migrations from outside of  the analysis phase space, which are due to the limited resolution of  
the $P_{\rm T,fwdjet}$ measurement. 

The subsample ``forward and central jet'' is selected by requiring an additional jet 
in the central region of the laboratory frame. This jet is required to have a transverse
momentum  $P_{\rm T, cenjet} > 4$ ${\rm  GeV}$ and to lie in the pseudorapidity region $-1 < \eta_{\rm cenjet} < 1$. The central jet must have a large rapidity separation from the most forward jet 
$\Delta \eta = (\eta_{\rm fwdjet} - \eta_{\rm cenjet}) > 2$. This condition enhances 
the phase space for additional parton emissions between the two jets. If there is more than one 
central jet, the one with the  smallest $\eta_{\rm cenjet}$ is chosen.

A summary of the selection cuts, defining the DIS phase space
for the measurement, the forward jet sample and the subsample with an additional central jet, 
is provided in table \ref{tab:disspace}.
With these requirements $13736$ and $8871$ events are selected for the forward jet and for the forward and central jet 
analysis, respectively. 
\renewcommand{\arraystretch}{1.15} 
\begin{table}[tb]
\begin{center}
\begin{tabular}{|c|c|c|}   \hline
{\bf DIS selection} & {\bf Forward jets}    &{\bf Central jets}\\
\hline
$0.1 < y < 0.7 $&$1.73~ <~\eta_{\rm{fwdjet}}~< 2.79$&
$-1 < \eta_{\rm{cenjet}}< 1$ \\[0.2cm]
$5 < Q^2 < 85$ ${\rm GeV}^2$&$P_{\rm{T,fwdjet}}~>~6$ ${\rm GeV}$&$P_{\rm{T,cenjet}}> 4$ ${\rm GeV}$  \\[0.2cm]
$0.0001< x < 0.004 $&$x_{\rm{fwdjet}} >0.035$ &
$\Delta \eta = \eta_{\rm{fwdjet}} -\eta_{\rm{cenjet}} > 2$\\[0.2cm]
&$0.5 < P^2_{\rm{T,fwdjet}}/Q^2 < 6$&\\[0.2cm]
\hline
\end{tabular}
\end{center}
\caption{Summary of cuts defining the DIS phase space, the forward jet and the central jet selection. 
If more than one forward jet is found, the jet with the largest $\eta_{\rm fwdjet}$ is chosen.
If there is more than one central jet, the one with the smallest $\eta_{\rm cenjet}$ is selected.}
\label{tab:disspace}
\end{table}
\subsection{Cross section determination}
In this measurement in addition to migrations between bins inside the measurement  
phase space, there are considerable  
migrations from outside of the analysis phase
space. This is taken into account in the calculation of the cross section corrected to the hadron level:  
\begin{equation}
\sigma_i = \frac{N^{\rm data}_i - N^{\rm out}_i }{\epsilon_i \cdot {\cal L}}.
\label{eq:02}
\end{equation}
Here  $N^{\rm data}_i$ is the number of observed events in bin $i$, $N^{\rm out}_i$ 
is the number of events from outside the measurement  phase space reconstructed 
in bin $i$, and $\epsilon_i$ is the efficiency in bin $i$. ${\cal L}$ is the total integrated luminosity.
$N^{\rm out}_i$  and $\epsilon_i$ are estimated using MC simulations. 
The purities\footnote{The purity is defined as the ratio of the number of events generated and reconstructed in the bin to the number of events originating from the phase space of the analysis and reconstructed in that bin.} 
in bins of the measured cross sections, as determined from the MC simulations, are at the level of $80\%$. 

The efficiency factors $\epsilon_i$ are calculated according to the formula :
\begin{equation}
\epsilon_i=\frac{N^{\rm det}_{i} - N^{\rm out}_i}{N^{\rm had}_i} ,
\label{eq:04}   
\end{equation}   
where $N^{\rm det}_{i}$ and $N^{\rm had}_i$ are the numbers of events in bin $i$ at the 
detector and  at the hadron level, respectively.
For this approach to be valid, the shape of  the distributions of all variables on which phase 
space cuts are applied have to be well described by the MC simulations also in the 
phase space extended beyond these cuts. This requirement is found to be satisfied by both 
models considered here.

The efficiency factors are calculated as the ratio of the model prediction 
at the detector level for a radiative MC
 and at the hadron level for a non-radiative MC, i.e. the data are also corrected 
for QED radiative effects. The efficiency factors are taken as the average of the factors estimated by the RAPGAP and DJANGOH/ARIADNE models. 
The uncertainty of the efficiency factors is taken to be half of the difference
between the factors
calculated using the two MC models and is included in the systematic error. 

\subsection{Systematic uncertainties}
The following sources of systematic uncertainties are considered :
\begin{itemize}
\item[-]
The model dependence of the bin-by-bin efficiency factors $\epsilon_i$ 
 leads to systematic uncertainties  between $2 \%$ and $6 \%$ for 
the measured cross sections.
\item[-]
The LAr hadronic energy scale uncertainty of $4 \%$ for this analysis gives rise to the dominant 
uncertainty of $7\%$ to $12 \%$ for the measured cross sections. 
\item[-]
The uncertainty on the electromagnetic energy scale of the SpaCal of $1 \%$ results in an uncertainty of the  measured cross sections below $3 \%$.
\item[-]
The uncertainty on the polar angle measurement of the scattered positron of 
$1$ $\rm mrad$ has 
a negligible effect on the cross section measurements.
\item[-]
The uncertainty on the determination of the trigger efficiency from the data, using independent 
 trigger samples, 
leads to an uncertainty between $2\%$ and $4\%$ on the cross section measurements.
\item[-]
The measurement of the integrated luminosity is accurate to within $1.5\%$.
\end{itemize}

The total systematic uncertainty, adding all individual contributions
quadratically, amounts to $11-12 \%$ for the measured cross sections.  

\section{Results}
\begin{figure}[hhh]
\center
\epsfig{file=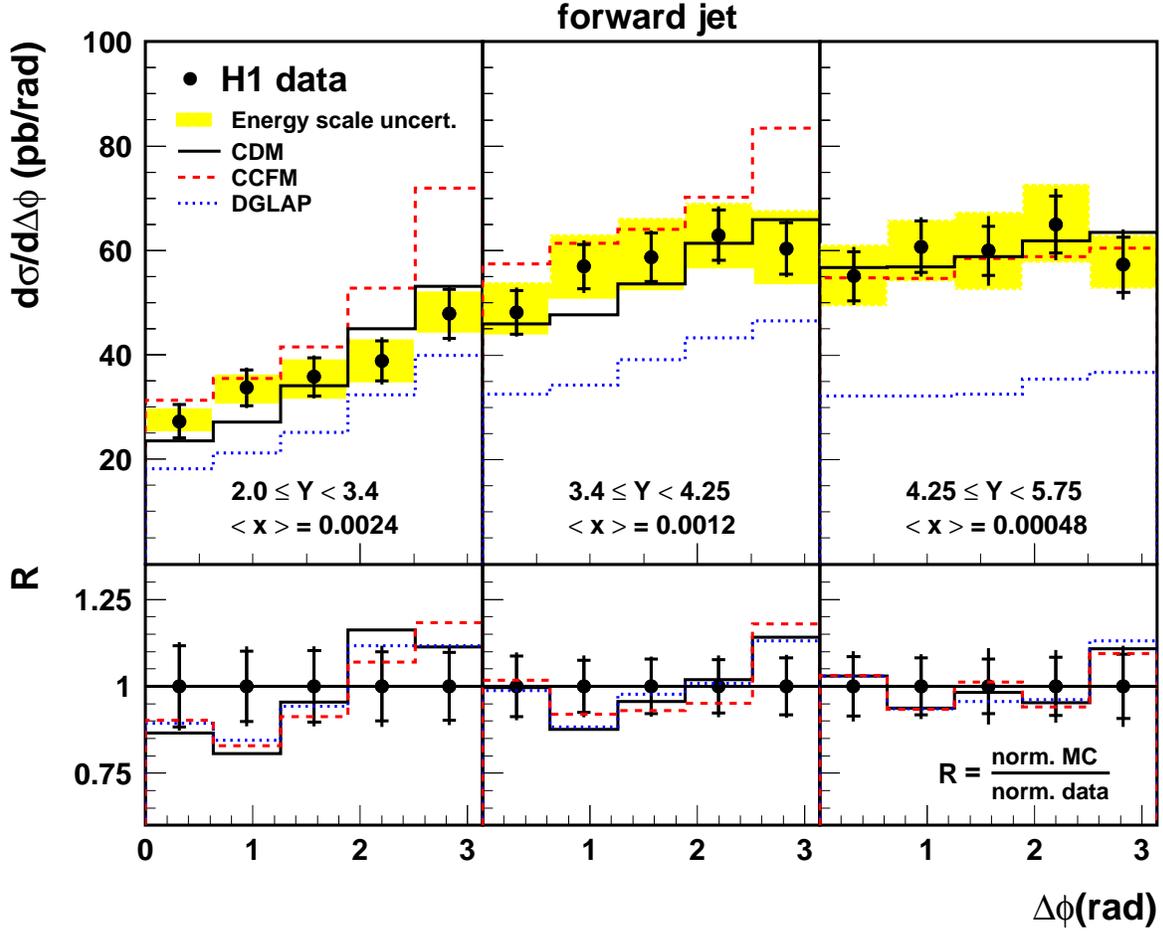 ,width=\textwidth}
\setlength{\unitlength}{1cm}
\caption{
Differential forward jet cross section as a function of the azimuthal angle 
difference $\Delta \phi$ 
between the most forward jet and scattered positron in three intervals 
of the variable $Y = \ln (x_{\rm fwdjet}/x)$. 
The inner error bars denote the statistical uncertainties. 
The systematic error due to the uncertainty of the hadronic energy scale  is shown separately 
as a band around the data points. Other systematic uncertainties  added quadratically to the statistical
uncertainties
are represented by the outer error bars.
The data are compared with the predictions of DJANGOH/ARIADNE (CDM) and RAPGAP (DGLAP)
with HERAPDF1.0, the CASCADE predictions (CCFM) are shown with uPDF set A0. 
In the lower part of 
the figure the ratio $R$ of MC to data for normalised cross sections is shown.
The precision of the measurements is shown at $R=1$ with the statistical and  
reduced systematic uncertainties indicated as error bars.    
}
\label{fig:azim1} 
\end{figure}

\begin{figure}[hhh]
\center
\epsfig{file=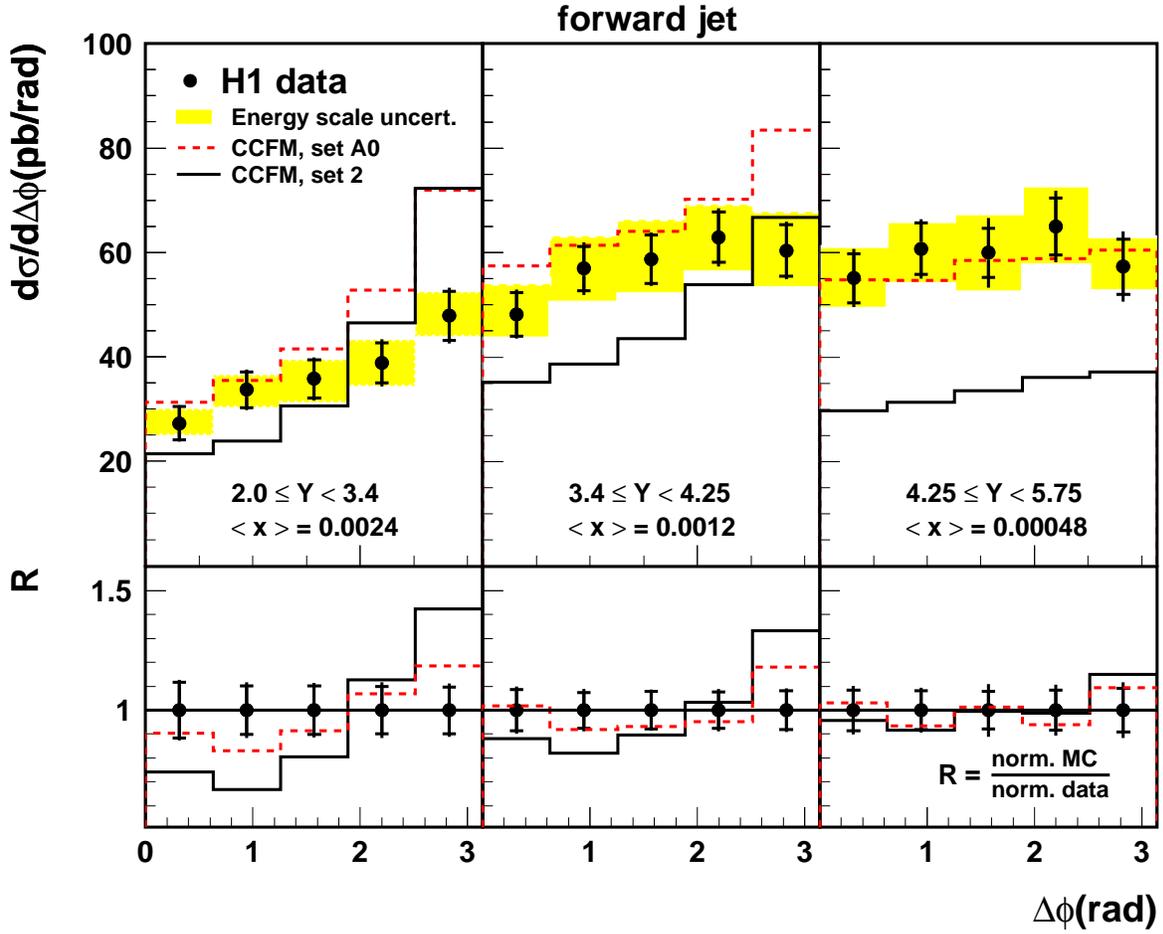 ,width=\textwidth}
\setlength{\unitlength}{1cm}
\caption{ 
Differential forward jet cross section as a function of the azimuthal angle difference $\Delta \phi$ 
between the most forward jet and the scattered positron in three intervals of 
the variable $Y = \ln (x_{\rm fwdjet}/x)$.
The data are compared to  the predictions of CASCADE (CCFM) with two different sets of  unintegrated 
gluon densities. For other details  see caption to figure 2. }
\label{fig:azim2} 
\end{figure}

\begin{figure}[hhh]
\center
\epsfig{file=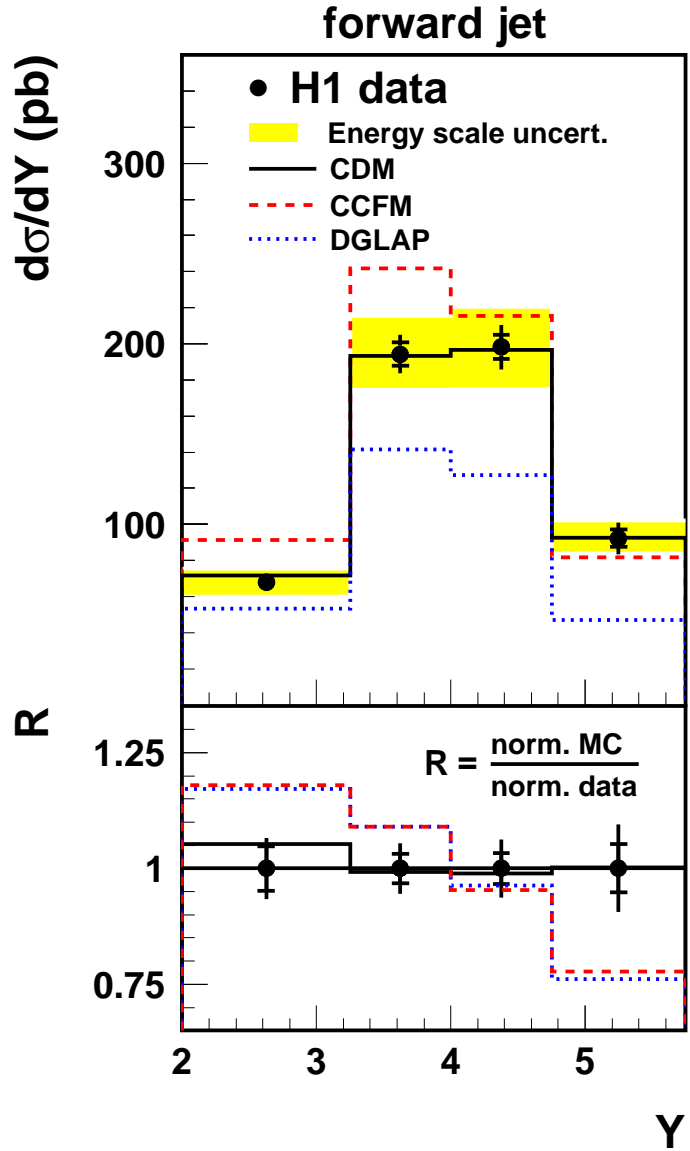 ,width=10cm}
\setlength{\unitlength}{1cm}
\caption{ Differential forward jet cross section as a function 
of the variable $Y = \ln (x_{\rm fwdjet}/x)$.
The data are compared with the predictions of DJANGOH/ARIADNE (CDM) and RAPGAP (DGLAP)
with HERAPDF1.0, the CASCADE predictions (CCFM) are shown with uPDF set A0. 
For other details see caption to figure 2.}
\label{fig:azim3} 
\end{figure}

The forward jet cross sections and their uncertainties  
are given in table~\ref{tab:mytest1} and presented in 
figures \ref{fig:azim1}-\ref{fig:azim3}.
Differential cross sections, $d\sigma/d\Delta \phi$, are presented as
a function of the azimuthal angle difference $\Delta \phi$ 
between the most forward jet and the scattered positron 
in bins of the variable $Y = \ln(x_{\rm fwdjet}/x)$.   
This variable approximates the rapidity distance  
between the scattered positron and the forward jet.
For the selected data 
sample the normalised shape distributions $1/ \sigma  \cdot d\sigma/d\Delta \phi$ are also 
determined, where $\sigma$ is
the integrated cross section in a given bin of $Y$. Furthermore, the forward jet cross 
section is measured as a function of $Y$.

The cross section $d\sigma/d\Delta\phi$ as a function of $\Delta \phi$  is shown in 
figure~\ref{fig:azim1} for three intervals of the variable 
$Y$: $2.0 \leq Y < 3.4$, $3.4 \leq Y < 4.25$ and $4.25 \le Y \le 5.75$. These $Y$ bins
correspond to average $x$ values of $0.0024$, $0.0012$ and $0.00048$, respectively.
At higher values of $Y$ 
the forward jet is more decorrelated from the scattered positron. 

The predictions of three QCD-based models with different underlying parton dynamics, discussed 
in section 2,  are compared with the data. 
The cross sections are well  described in shape and normalisation by CDM which has a
 BFKL-like approach.
Predictions of RAPGAP, which implements DGLAP evolution, fall below the data, 
particularly at large $Y$. 
Calculations in the CCFM scheme as implemented in CASCADE using the uPDF set A0~\cite{setA0}
 overestimate the measured cross section for large $\Delta \phi$ values in the two lowest 
$Y$ intervals. However, this model provides as good a description as CDM 
of the data in the highest $Y$ interval. 

The shape of the $\Delta \phi$ distributions,  $1 / \sigma \cdot d\sigma/d\Delta \phi$, is compared to the different MC predictions in the lower part
of figure~\ref{fig:azim1}, where the ratio $R$ is shown, defined as:
\begin{equation}
R = \left(\frac{1}{\sigma^{\rm MC}} \frac{d\sigma^{\rm MC}}{d\Delta \phi} \right) \, \Big/ \, \left(\frac{1}{\sigma^{\rm data}} \frac{d\sigma^{\rm data}}{d\Delta \phi}\right) \, .
\label{eq:1}
\end{equation}
The precision of the measurements is shown at $R=1$ where the statistical and 
systematic uncertainties are indicated.    
The systematic uncertainty is reduced in the ratio and contains only two 
components added in quadrature: the model dependence of 
the correction factors and the trigger efficiency uncertainty. 
The ratio plots show that in the analysed phase space region the 
shape of the $\Delta \phi$ distributions is well described by all MC models. 
Since the shape predictions of the three models are very similar, this observable alone
cannot discriminate among the models.  
It should be noted that  the shape of the $\Delta \phi$ distributions 
is rather insensitive to the PDF used for event generation. The shape distributions generated using 
CTEQ6L, CTEQ6M \cite{CTEQ6L} and HERAPDF1.0 \cite{HERAPDF} differ on average by $1$-$2\%$.
However, the cross section normalisation is more sensitive to the choice of PDF with 
differences up to $5\%$ for CDM and up to $20\%$ for RAPGAP at large $Y$. 

Predictions of the CCFM model presented in figure~\ref{fig:azim2} indicate a significant 
sensitivity to the choice of the uPDF. Set A0 and J2003-set 2 give quite different predictions 
for the differential cross sections in all $Y$ intervals. 
Set A0 provides a reasonable description of the measured cross sections, except for the 
region of large $\Delta \phi$ in the two lowest $Y$ bins. Predictions using J2003-set 2 do 
not describe the data, especially at higher $Y$, where the estimated  cross sections are too low.
 The shape of the $\Delta \phi$ distributions is reasonably well described by 
 the set A0. At low $Y$ 
it shows sensitivity to the unintegrated gluon density.

The cross section $d\sigma/dY$ as a function of the rapidity separation $Y$  
is shown in  figure~\ref{fig:azim3}. 
The CDM model describes the data well 
over the whole $Y$ range.
The DGLAP predictions  fall below the data, but approach them at small $Y$. 
The predictions of  the CCFM model  are above the data at small $Y$ but describe them
well at larger $Y$ corresponding to low values of $x$.

\begin{figure}[hhh]
\center
\epsfig{file=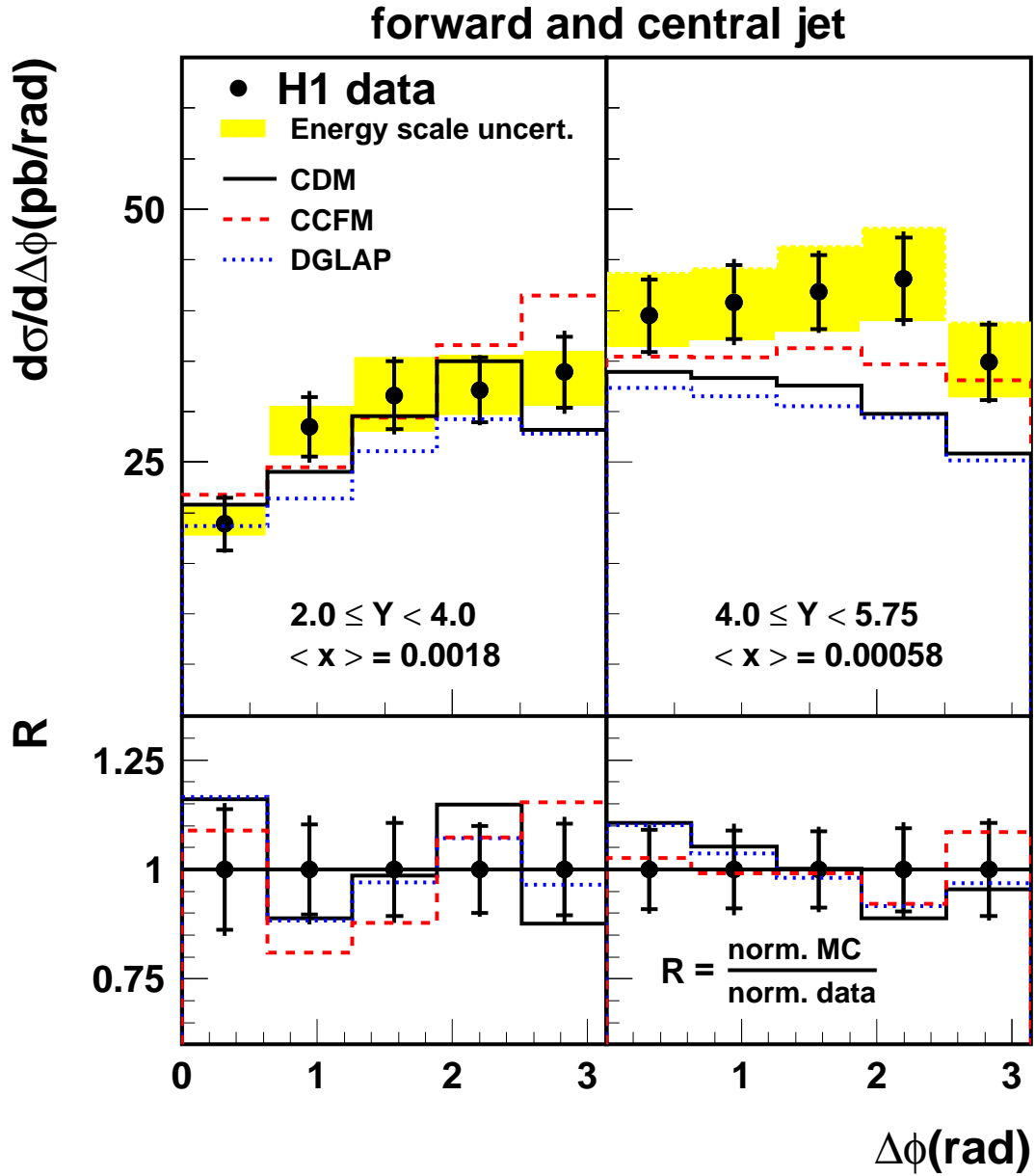 ,width=15cm}
\setlength{\unitlength}{1cm}
\caption{ 
Differential forward and central jet cross section as a function of the azimuthal angle difference $\Delta \phi$ 
between the most forward jet and the scattered positron 
in two intervals of the variable $Y = \ln (x_{\rm fwdjet}/x)$.
The data are compared with the predictions of DJANGOH/ARIADNE (CDM) and RAPGAP (DGLAP)
with HERAPDF1.0, the CASCADE predictions (CCFM) are shown with uPDF set A0. 
For other details see caption to figure 2.
}
\label{fig:azim4} 
\end{figure}

The forward and central jet cross sections and their uncertainties are given in table~\ref{tab:mytest2}.
The differential cross section $d\sigma/d\Delta\phi$ as a function of the azimuthal angle 
difference $\Delta \phi$  is shown in figure~\ref{fig:azim4} in comparison with 
the predictions of the three MC models.
The cross sections are measured in two intervals of $Y$, 
$2.0 \le Y < 4.0$  and $4.0 \le Y \le 5.75$.

From figure~\ref{fig:azim4} it is observed that at lower $Y$ the predictions of all models 
describe the cross sections reasonably well.  
At high $Y$ all models undershoot the data: CCFM  (set A0) is closest to the data, 
the DGLAP and  CDM predictions are below the measured cross section. 
The ratio $R$ in the lower part of figure~\ref{fig:azim4} shows that the shape of the 
$\Delta \phi$ 
distributions is well described by all MC models, as in the case of the forward jet measurements. 

\begin{figure}[hhh]
\center
\epsfig{file=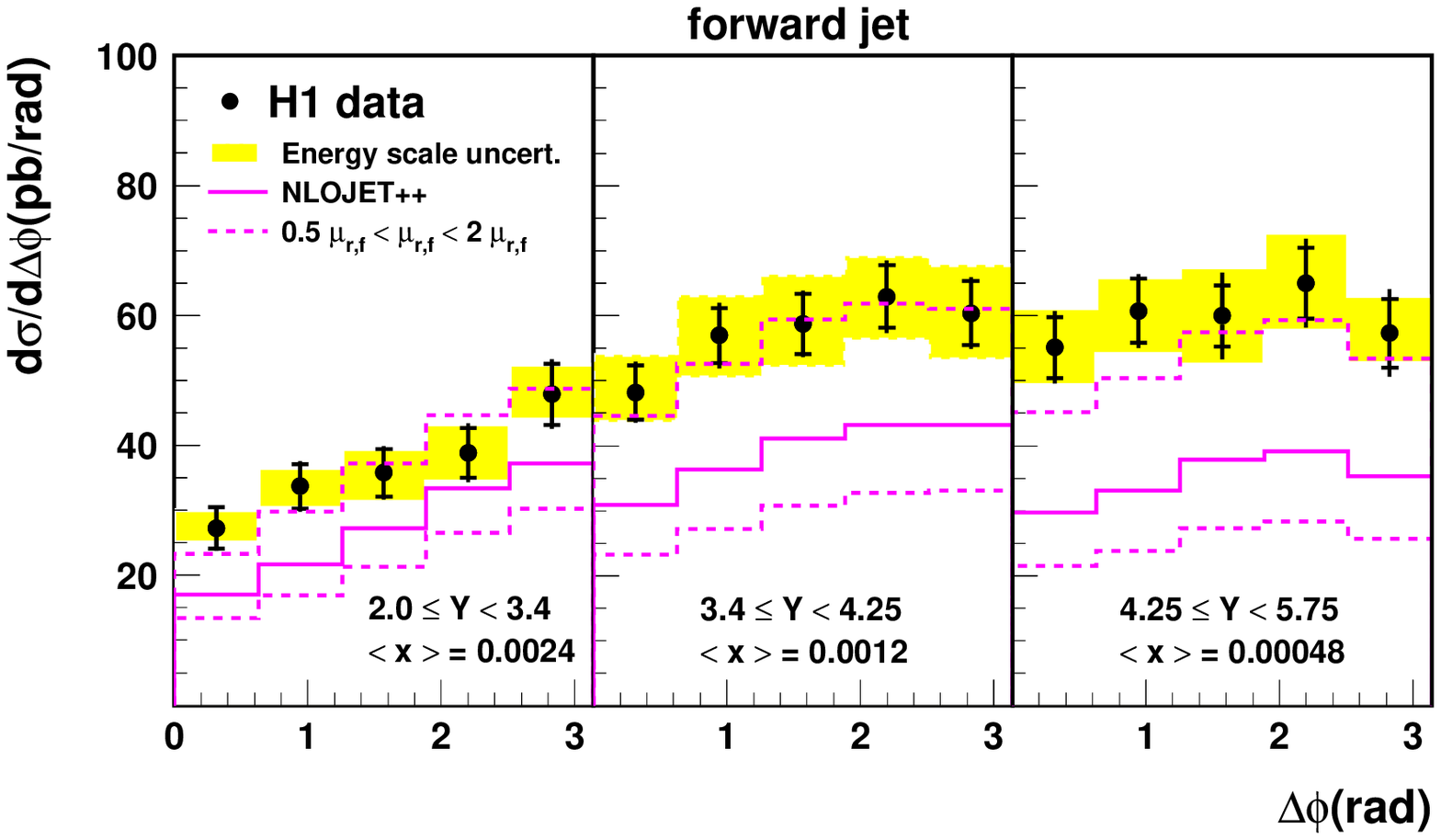 ,width=15cm}
\setlength{\unitlength}{1cm}
\caption{ 
Differential forward jet cross section as a function of the azimuthal angle difference $\Delta \phi$ 
between the most forward jet and the scattered positron 
in three intervals of the variable $Y = \ln (x_{\rm fwdjet}/x)$.
The data are compared to the corrected to hadron level NLO predictions from NLOJET++ which uses the CTEQ6.6 PDF.
Dashed lines above and 
below the nominal NLO prediction show theoretical uncertainty estimated by applying a factor $2$
or $1/2$ to the renormalisation and factorisation scales simultaneously. For other details see caption
to figure 2.}

\label{fig:nloinc} 
\end{figure}

\begin{figure}[hhh]
\center
\epsfig{file=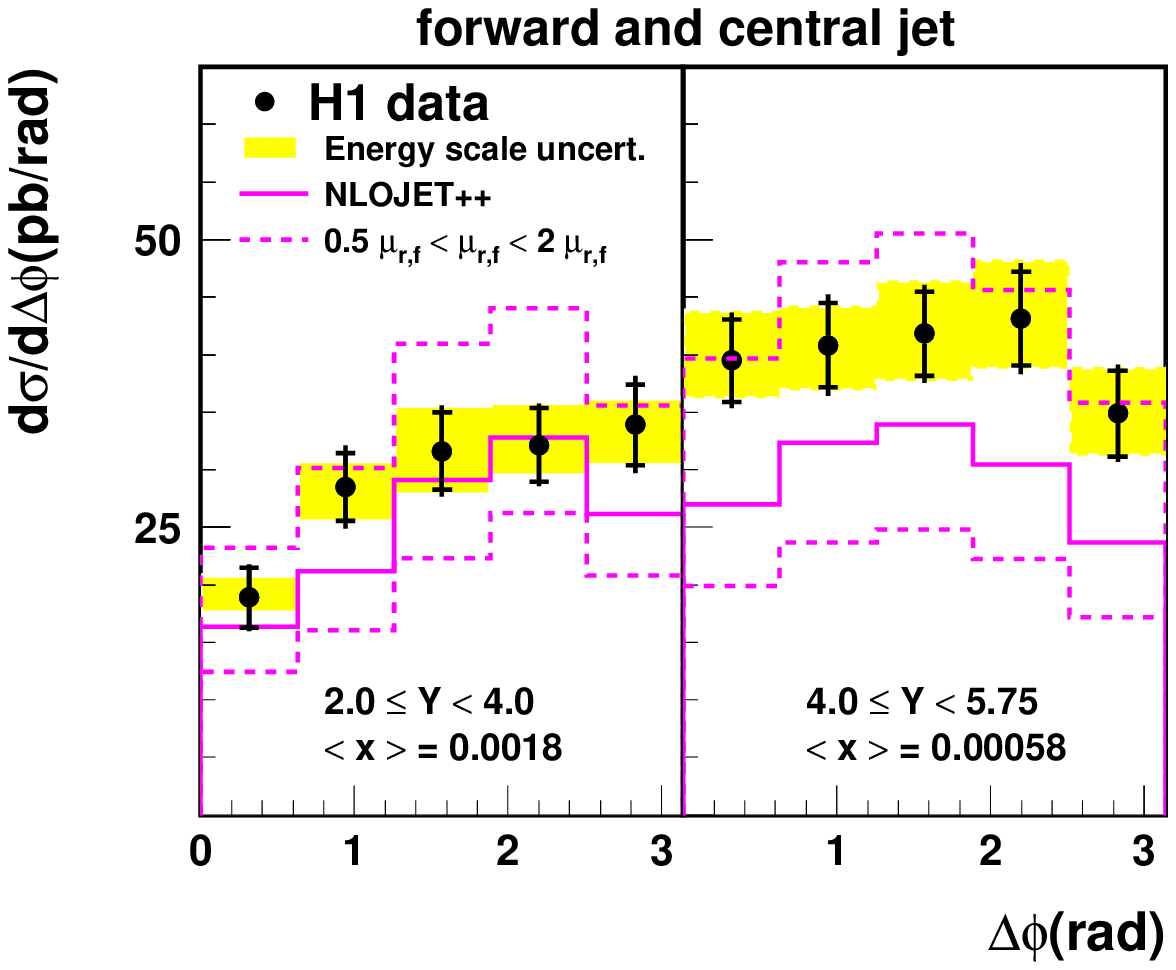 ,width=15cm}
\setlength{\unitlength}{1cm}
\caption{ 
Differential forward and central jet cross section as a function of the azimuthal angle difference $\Delta \phi$ 
between the most forward jet and the scattered positron in two intervals 
of the variable $Y = \ln (x_{\rm fwdjet}/x)$.
The data are compared to the corrected to hadron level NLO predictions from NLOJET++ which uses the CTEQ6.6 PDF. 
Dashed lines above and 
below the nominal NLO prediction show theoretical uncertainty estimated by applying a factor $2$
or $1/2$ to the renormalisation and factorisation scales simultaneously. For other details see caption
to figure 2.}

\label{fig:nlo} 
\end{figure}

Comparisons of the measured $\Delta \phi$ distributions 
with NLOJET++ predictions are shown in figures \ref{fig:nloinc} and \ref{fig:nlo}.
 The calculations are performed at $O(\alpha_S^2)$ precision using  
the CTEQ6.6 PDF \cite{CTEQ6.6} and 
$\alpha_S(M_Z) = 0.118$. 
Large theoretical uncertainties of up to $50\%$ from the variation of  factorisation and 
renormalisation scales are observed. 
The size of the theoretical uncertainty indicates that in this phase space region 
higher order contributions are expected to be important. 

In the forward jet sample (figure \ref{fig:nloinc}) 
for all three ranges of $Y$ the data are 
above the central NLO result but still within the theoretical uncertainty. 
In the case of the forward and central jet sample 
shown in figure \ref{fig:nlo}, the  NLO calculation describes the 
data at low $Y$. Only at high $Y$ in the regime of the BFKL evolution it is below the data, but again within the large 
theoretical uncertainty.

In summary,
the correlation between the forward jet and the positron decreases with $Y$ and 
the $\Delta \phi$ distributions are flat at high $Y$.
The measurements of the forward jet cross sections favour CDM and disfavour the RAPGAP model.
CASCADE provides a reasonable description of the data at large $Y$, but shows sizeable sensitivity
to the uPDF.
The shape of the measured $\Delta \phi$ distributions is well described by MC models based on different
QCD evolution schemes.

The similarity of the $\Delta \phi$ shapes of the MC predictions suggests that the forward jet predominantly
originates from the hard matrix elements which are similar in all three models. 
However, MC studies with RAPGAP show 
that $80\%$ of the forward jets are produced by  parton showers. When the initial state parton
shower is switched off, the shape of the $\Delta \phi$ distribution is only
slightly changed, but the
normalisation is significantly reduced. 
This indicates that the decorrelation in $\Delta \phi$
is mainly governed by the phase space requirements, in particular by the 
rapidity separation $Y$,
and that the normalisation of the cross sections is mainly influenced by the 
amount of soft radiation from
parton showers, which depends on the evolution scheme. 

\section{Conclusions} 
Measurements of DIS events at low $Q^2$ containing a high transverse momentum jet produced
 in the forward direction, at small angles with respect to the  proton beam, are presented.
Differential cross sections and normalised distributions are measured as a function of the azimuthal angle difference $\Delta \phi$ and the rapidity separation $Y$ between the forward jet and the scattered positron.
Investigations of the azimuthal correlation between the most forward jet and the outgoing positron  
are performed in different regions of  $Y$ for the forward jet sample
and for the subsample  with an additional  central jet. To test the sensitivity of the measured observables to QCD dynamics at low $x$, the data are compared to QCD models with different parton evolution approaches and to predictions 
of next-to-leading order QCD calculations.

Measurements of the cross sections as a function of $\Delta \phi$ and $Y$ are best described by the BFKL-like CDM model, while 
the DGLAP-based RAPGAP model is substantially below the data.
The CCFM-based CASCADE provides a reasonable description of the data but shows sizeable sensitivity to the unintegrated
gluon density. The shape of the $\Delta \phi$ distributions does not 
discriminate further between different evolution schemes.
The fixed order NLO DGLAP predictions are in general below the data, but still
in agreement within the large theoretical uncertainties.  

\section*{Acknowledgements}

We are grateful to the HERA machine group whose outstanding efforts have made this experiment possible. We thank the engineers and technicians for their work in constructing and maintaining the H1 detector, our funding agencies for financial support, the DESY technical staff for continual assistance and the DESY directorate for support and for the hospitality which they extend to the 
non-DESY members of the collaboration.


\clearpage
\newpage
\renewcommand{\arraystretch}{1.30} 
\begin{table}[tb]
  \begin{center}
    \begin{tabular}{|c|c|c|c|c|}
      \hline
      {\boldmath $ \Delta \phi$ \bf range [rad]} & {\boldmath $ d\sigma / d\Delta \phi$ \bf [pb/rad] }&{\boldmath ${\delta_{stat}}$ \bf [pb/rad]} &{\boldmath $\delta_{had}$ \bf [pb/rad] } & {\boldmath $\delta_{syst}$ \bf [pb/rad]}  \\
      \hline 
   \multicolumn{5}{|c|} {$2.0 \le Y < 3.4 $} \\
      \hline
       $0.0\phantom{0}-0.63$ & $27.3$  & $\pm3.2$   & $^{+2.7}_{-2.2}$  & $^{+1.3}_{-1.3}$ \\
       $0.63-1.26$   & $33.7$  & $\pm3.4$   & $^{+2.8}_{-3.3}$  & $^{+1.9}_{-1.8}$ \\
       $1.26-1.89$   & $35.8$  & $\pm3.7$   & $^{+3.6}_{-4.4}$  & $^{+2.2}_{-1.8}$ \\
       $1.89-2.51$   & $38.9$  & $\pm3.8$   & $^{+4.4}_{-4.4}$  & $^{+2.3}_{-2.6}$ \\
       $2.51-3.14$   & $47.9$  & $\pm4.7$   & $^{+4.6}_{-3.9}$  & $^{+2.7}_{-2.5}$ \\
      \hline
   \multicolumn{5}{|c|}{ $3.4 \le Y < 4.25 $} \\
      \hline  
       $0.0\phantom{0}-0.63$ &  $48.2$  & $\pm4.2$ & $^{+5.8}_{-4.5}$  & $^{+2.1}_{-2.3}$ \\
       $0.63-1.26$   &  $56.9$  & $\pm4.3$ & $^{+6.1}_{-6.4}$  & $^{+2.8}_{-2.6}$ \\
       $1.26-1.89$   &  $58.7$  & $\pm4.6$ & $^{+7.5}_{-6.6}$  & $^{+2.3}_{-2.1}$ \\
       $1.89-2.51$   &  $62.9$  & $\pm4.8$ & $^{+6.3}_{-6.6}$  & $^{+2.6}_{-3.1}$ \\
       $2.51-3.14$   &  $60.4$  & $\pm4.9$ & $^{+7.3}_{-7.1}$  & $^{+2.3}_{-2.7}$ \\
      \hline
   \multicolumn{5}{|c|}{ $4.25 \le Y < 5.75 $} \\
      \hline
       $0.0\phantom{0}-0.63$ &  $55.1$  & $\pm4.7$ & $^{+6.0}_{-5.8}$ & $^{+2.9}_{-3.2}$ \\
       $0.63-1.26$   &  $60.8$  & $\pm5.0$ & $^{+5.2}_{-6.7}$ & $^{+2.6}_{-2.9}$ \\
       $1.26-1.89$   &  $60.0$  & $\pm4.7$ & $^{+7.4}_{-7.4}$ & $^{+4.6}_{-4.8}$ \\
       $1.89-2.51$   &  $65.0$  & $\pm5.4$ & $^{+7.7}_{-7.3}$ & $^{+4.2}_{-4.1}$ \\
       $2.51-3.14$   &  $57.3$  & $\pm5.3$ & $^{+5.7}_{-4.6}$ & $^{+4.2}_{-4.1}$ \\
      \hline
      \hline
      {\boldmath $Y$ \bf range  }& {\boldmath $ d\sigma / dY $ \bf [pb]}&{\boldmath $\delta_{stat}$ \bf [pb]} &{\boldmath $\delta_{had}$ \bf [pb]} &{\boldmath  $ \delta_{syst}$ \bf [pb]}  \\
      \hline
       $2.00-3.25$      &  $\phantom{0}67.9$ & $\pm3.3$ & $~\,^{+7.5}_{-7.7}$ & $~\,^{+3.0}_{-3.1}$ \\
       $3.25-4.00$      &  $194.4$   & $\pm6.3$ & $^{+21.1}_{-20.0}$  & $~\,^{+8.3}_{-8.7}$ \\
       $4.00-4.75$      &  $198.2$   & $\pm6.7$ & $^{+22.5}_{-23.4}$  & $^{+10.4}_{-10.4}$ \\
       $4.75-5.75$      &  $\phantom{0}92.3$ & $\pm4.8$ & $~\,^{+9.7}_{-8.3}$ & $~\,^{+7.2}_{-7.3}$ \\
      \hline

    \end{tabular}
    \caption{Differential forward jet cross section in bins of 
the variable $Y = \ln (x_{\rm fwdjet}/x)$ 
and the azimuthal angle difference $\Delta \phi$ between the most forward jet 
and the scattered positron. 
The statistical uncertainty ($\delta_{stat}$), the uncertainty due to the hadronic energy 
scale ($\delta_{had}$) and other systematic uncertainties ($\delta_{syst}$) described in  
the text are given.}  
    \label{tab:mytest1}
  \end{center}
\end{table}
\clearpage
\newpage
\renewcommand{\arraystretch}{1.30} 
\begin{table}[tb]
  \begin{center}
    \begin{tabular}{|c|c|c|c|c|}
      \hline
      {\boldmath $\Delta \phi$ \bf range [rad]}& {\boldmath $ d\sigma / d\Delta \phi$ \bf [pb/rad] }&{\boldmath $\delta_{stat}$ \bf [pb/rad]} & {\boldmath $\delta_{had}$ \bf [pb/rad] }& {\boldmath $ \delta_{syst}$ \bf [pb/rad]}  \\
      \hline 
   \multicolumn{5}{|c|}{ $2.0 \le Y < 4.0 $} \\
      \hline
       $0.0\phantom{0}-0.63$ & $18.9$  & $\pm2.6$   & $^{+1.9}_{-1.3}$  & $^{+1.3}_{-1.1}$ \\
       $0.63-1.26$   & $28.5$  & $\pm2.9$   & $^{+2.3}_{-2.9}$  & $^{+2.0}_{-1.9}$ \\
       $1.26-1.89$   & $31.6$  & $\pm3.4$   & $^{+3.9}_{-3.8}$  & $^{+1.9}_{-1.9}$ \\
       $1.89-2.51$   & $32.1$  & $\pm3.2$   & $^{+3.6}_{-2.7}$  & $^{+1.3}_{-1.3}$ \\
       $2.51-3.14$   & $33.9$  & $\pm3.5$   & $^{+2.3}_{-3.4}$  & $^{+2.1}_{-2.1}$ \\
      \hline
   \multicolumn{5}{|c|}{ $4.0 \le Y < 5.75 $} \\
      \hline  
       $0.0\phantom{0}-0.63$ &  $39.5$  & $\pm3.6$ & $^{+4.3}_{-3.3}$  & $^{+1.6}_{-1.9}$ \\
       $0.63-1.26$   &  $40.8$  & $\pm3.6$ & $^{+3.4}_{-3.9}$  & $^{+2.2}_{-2.2}$ \\
       $1.26-1.89$   &  $41.8$  & $\pm3.7$ & $^{+4.6}_{-4.1}$  & $^{+1.8}_{-1.8}$ \\
       $1.89-2.51$   &  $43.1$  & $\pm4.1$ & $^{+5.2}_{-4.4}$  & $^{+2.2}_{-2.2}$ \\
       $2.51-3.14$   &  $34.9$  & $\pm3.7$ & $^{+4.0}_{-3.7}$  & $^{+1.7}_{-1.7}$ \\
      \hline
    \end{tabular}
    \caption{Differential forward and central jet cross section in bins of 
of the variable $Y = \ln (x_{\rm fwdjet}/x)$ and
the azimuthal angle difference $\Delta \phi$  
between the most forward jet and the scattered positron.  
The statistical uncertainty ($\delta_{stat}$), the uncertainty 
due to the hadronic energy scale ($\delta_{had}$), and other systematic 
uncertainties ($\delta_{syst}$) described in the text are given.}  
    \label{tab:mytest2}
  \end{center}
\end{table}
\clearpage
\newpage
\newpage
\newpage
\newpage
\newpage
\newpage

\end{document}